\documentstyle[pra,epsf,aps,twocolumn]{revtex}
\begin{document}
\sloppy
\title{Functional determinants via Wronski construction
    of Green functions}
\author{H.\ Kleinert\thanks{E-mail: kleinert@physik.fu-berlin.de} and
     A.\ Chervyakov\thanks{On leave from LCTA, JINR, Dubna, Russia}
              \\
         Freie Universit\"at Berlin\\
          Institut f\"ur Theoretische Physik\\
          Arnimallee14, D-14195 Berlin
     }
\maketitle
\begin{abstract}
A general technique is developed for calculating functional determinants
of second-order differential operators
with Dirichlet, periodic, and antiperiodic boundary conditions.
As an example, we give simple formulas
 for a
  harmonic oscillator with an arbitrary
time-dependent frequency.
Here our result is a generalization of Gel'fand-Yaglom's
famous formula which was restricted to Dirichlet boundary conditions.
Apart from the generalization,
our derivation
is more transparent than theirs,
 the determinants requiring only
knowledge
of the classical trajectories.
Special properties of operators
 with a zero mode are exhibited.
Our technique does not require
the calculation of the spectrum and is
as simple as Wronski's method for
Green functions.
\end{abstract}
1.\  Evaluation of  Gaussian path integrals
 is necessary in many physical problems.
In particular, it  appears
in all semi-classical calculations
of fluctuating systems.
Typically, it leads to the problem
of calculating
 the
 functional determinant of a second-order differential
operator \cite{1}.
For Dirichlet boundary conditions,
a first general solution of this problem was given by Gel'fand and Yaglom
\cite{2} based on the lattice approximation  to
 path integrals in the continuum limit.
 Their
  result  was  expressed
in terms of a
  simple differential equation
for the functional determinant.
In subsequent work \cite{3}--\cite{5}, the formalism was
generalized to a variety of differential operators and boundary
conditions based on the concept of zeta-function
regularization \cite{7}.
%In subsequent work \cite{3}--\cite{5}, the formalism was
% generalized to a variety of linear operators and boundary
%conditions.
Unfortunately,  Gel'fand-Yaglom's method becomes  rather
complicated
 for  periodic and antiperiodic boundary
condition relevant in  quantum statistic
(see Section 2.12 in \cite{1}).
In the periodic case there is, moreover,  a zero
mode
causing additional complications.

In this paper we present a systematic method for
finding
functional determinants of linear differential operators
which is based on  Wronski's simple
construction of Green functions.
Our method is simpler than those used
in the
previous
approaches,
 since the
determinants are expressed
entirely in terms of a classical trajectory. Furthermore,
for fluctuation operator with a zero
mode, a case frequently encountered in  semiclassical calculations,
 the  special treatment of this mode  becomes transparent.
\\~\\
2.\
The typical  fluctuation action
arising in
 semiclassical approximations
has a quadratic Lagrangian of the form
\begin{equation}
  L = \frac{M}{2} \left[ \dot x^2 -  \Omega ^2 (t) x^2 \right].
\label{1}\end{equation}
 Physically, this Lagrangian  describes a  harmonic oscillator with a
time-dependent frequency $ \Omega (t)$. The path integral
 for such a system was studied
in  several papers \cite{cheng}--\cite{8}. For such an oscillator, both
the quantum mechanical propagator and the thermal partition
 function contain a phase factor $\exp [i S_{\rm cl} (x)]$
and are multiplied by a pre-exponential
 fluctuation factor  proportional to
\begin{equation}
  F(t_b, t_a) \sim  \left( \frac{{\rm Det} K_1}{{\rm Det} \tilde K}
\right)^{-1/2},
\label{2}\end{equation}
where $K_1 = -\partial^2_t -  \Omega ^2 (t) \equiv K_0 -  \Omega ^2 (t)$
is the kernel of the second variation of action $S(x)$ along
the classical path $x_{\rm cl}(t)$. The linear operator $K_1$ acts on the space
 of twice differentiable functions $y(t) =  \delta x (t)$ on an interval
$t \in [t_a, t_b] $ with  appropriate boundary conditions.
These are Dirichlet boundary conditions $y(t_a) = y (t_b) = 0$
in the quantum-mechanical case, and periodic (antiperiodic)
$y(t_b) = \pm y (t_a), \dot y (t_b) = \pm \dot y (t_a)$
in the quantum statistical case. In these two cases the
operator $\tilde K$ may be chosen as $ K_0$
 or $K_0 -  \omega _0^2 $,
respectively, where $ \omega _0 $ is a time-independent
 oscillator frequency. The ratio of determinants (\ref{2})
 arises naturally
from the normalization of the path integral
and is well-defined \cite{1}.
%%%%
Furthermore, for such an operator we may assume
the Fredholm property
\begin{equation}
 \frac{{\rm Det}K_1}{{\rm Det}\tilde K} =
   {\rm Det} \tilde K^{-1} K_1
\label{3a}\end{equation}
thus neglecting multiplicative anomalies \cite{12}.
Since the operator $\tilde{K}^{-1}K_1$ is of the form $I + B$,
 with $B$ an operator of the trace class, it has a well-defined
 determinant even without any  regularization.

To calculate $F(t_b,t_a)$, we introduce
 a one-parameter family of
operators $K_g$ depending linearly on the parameter $g: K_g = K_0 -
 g  \Omega ^2 (t),
 0 \ll g \ll 1$.
The above property (\ref{3a}) allows us to
 make use of the well-known formula
$\log {\rm Det} \tilde K^{-1} K_g = {{\rm Tr}} \log \tilde  K ^{-1} K_g$ to
relate the $g$-derivative of the logarithm of the ratio (\ref{2}) to the
trace of the Green function  of the operator $K_g$ as follows
\begin{equation}
  \partial_g \log {\rm Det} \tilde K^{-1} K_g
= -{\rm Tr}~  \Omega ^2 (t) G_g (t,t'),
\label{3}\end{equation}
 the Green function being defined by
\begin{equation}
  G_g (t,t') = [- \partial^2_t - g  \Omega ^2 (t)]^{-1}  \delta (t-t').
\label{4}\end{equation}
Formula (\ref{3}) is valid provided
we
regularize
the trace
on the right-hand side, if it diverges,
via
  zeta-functions  $\zeta (s) = \sum_{i}  \lambda ^{-s}$, where the sum
runs over all
eigenvalues. It is convergent for sufficinetly large $s$
and defined for smaller $s$ by analytic continuation
(see  \cite{7}).
Then,
 for each member of the $g$-family,
${\rm Det} K_1 = \exp [- \xi' (0)]$.
Another proof of (\ref{3}) can be found in \cite{3}.

By integrating
 (\ref{3}), we obtain  for the ratio of functional determinants
(\ref{3a}):
\begin{equation}
{\rm Det} \tilde K^{-1} K_g = C \exp \left\{- \int_{0}^{g} dg'
       \int^{t_b}_{t_a} dt\,  \Omega ^2 (t) G_{g'} (t,t) \right\} ,
\label{5}\end{equation}
where $C = {\rm Det} \tilde K^{-1} K_0$ is a $g$-independent
constant.
This is our basic formula to be
 supplemented
 by an appropriate boundary condition
 to Eq.~(\ref{4}) for the Green function
as we shall now discuss in detail.
\\~\\
3.
A general solution of Eq.~(\ref{4}) is given by
advanced or retarded Green functions as follows
\begin{equation}
 G_g^- (t,t') = G^+_g (t',t) =  \Theta_{tt'} \cdot f_g
     (t,t'),
\label{7a}\end{equation}
where $\Theta_{tt'} = \Theta (t-t')$ is Heaviside's function
and $f_g(t,t')$ is a combination
\begin{equation}
 f_g(t,t') = \frac{1}{W_g} \left[ \eta_g (t) \xi_g(t') -
   \xi_g (t)\eta _g(t')\right]
\label{8a}\end{equation}
 of
two linearly independent
 solutions
$\eta_g (t)$ and $\xi_g (t)$
of the homogeneous equation
\begin{equation}
  \left[ - \partial^2_t - g  \Omega ^2 (t)\right]  h_g(t) = 0.
\label{9a}\end{equation}
The constant $W_g$ is the time independent Wronski determinant
 $W _g= \eta_g \dot \xi_g - \dot \eta_g \xi_g $. The solution
(\ref{7a}) is not unique since it leaves  room
 for an additional general solution of the homogeneous
equation (\ref{9a}) with an arbitrary coefficients.
 This freedom is
 removed
 by  appropriate boundary conditions.
Consider first the
quantum mechanical case which requires
Dirichlet boundary conditions
$y (t_b) = y (t_a) = 0$
for the eigenfunctions $y(t)$
of $K_1$,
implying for the Green
function the boundary conditions
\begin{eqnarray}
   G_g(t_a, t') & = & 0,~~~t \leq t'. \nonumber\\
   G_g(t',t_b) & = & 0,~~~t'\leq t.
\label{6}\end{eqnarray}
The operator
$\tilde K$  in the ratio
(\ref{2})  is  equal to
 $ K_0$,  and the constant C in Eq.~(\ref{5}) is
 unity. After imposing
  (\ref{6}), the Green function
is uniquely  given by  Wronski's formula:
\begin{equation}
 G_g (t,t') = \frac{\Theta_{tt'} f_g (t',t_a) f_g (t_b,t)
 + \Theta_{t't} f_g (t,t_a) f_g (t_b,t') }{f_g (t_a, t_b)},
\label{11a}\end{equation}
where
\begin{equation}
 f_g (t_a, t_b) = \frac{{\rm Det}  \Lambda_g }{W_g} \neq 0,
\label{12a}\end{equation}
  with
 $\Lambda$  being  a constant $(2 \times 2)$-matrix
\begin{equation}
\Lambda = \left(
\begin{array}{ll}
      \eta_a & \xi_a\\
    \eta_b & \xi_b
\end{array}\right),
\label{13a}\end{equation}
 formed from the solutions
$\eta_g(t)$ and $\xi_g (t)$
at arbitrary
$g\neq 1$.
Note that these solutions
are restricted only the condition (\ref{12a}).
The result is unique and well-defined, assuming the absence of
 a zero mode $\xi (t)$ of the
operator $K_1$
with Dirichlet boundary  conditions
$\xi_a = \xi_b = 0$. Such a mode
would cause problems
since according to
(\ref{6}),
the Wronski determinant
$W$ would  vanish  at the initial point,
and thus for all $t$.

Excluding
zero  modes, we obtain from
(\ref{8a}):
\begin{equation}
 {\rm Tr} \, \Omega ^2 (t) G_g (t,t') \!=\! \frac{1}{f_g (t_a, t_b)}
  \! \int^{t_b}_{t_a} dt\,  \Omega ^2 (t) f_g (t,t_a) f_g (t_b,t).
\label{14a}\end{equation}
 To perform the time integral on the right hand side,
  we  make use of the identity
\begin{equation}
\Omega ^2 (t) \xi (t)\eta (t)
 = \partial_t [\dot \eta_g(t) \partial_g \xi_g(t) -
     \eta_g(t) \partial_g \dot \xi_g(t)].
\label{10}\end{equation}
 This follows from Eq.~(\ref{9a})  for $\eta (t,g)$,
and an analogous equation
 for $\xi_{\tilde g} (t) $, after  multiplying the first by
 $\xi_{\tilde g} (t) $ and the second by $\eta_g (t)$,
 and  taking their difference. In the limit
 $\tilde g \rightarrow g   $, we  obtain (\ref{10})
 from the linear term  in $\tilde g - g$.
Inserting (\ref{10}) into (\ref{14a}), we see that
\begin{equation}
{\rm Tr} \, \Omega ^2 (t) G_g (t,t') = -\partial_g \log
   \left(\frac{{\rm Det}  \Lambda_g}{W_g}\right).
\label{16a}\end{equation}
  Substituting (\ref{16a}) into (\ref{5}),
   we find
\begin{equation}
 {\rm Det}  K_0^{-1} K_g = \frac{{\rm Det} \Lambda_g}{W_g}\bigg/
    \frac{{\rm Det}  \Lambda _0}{W_0},
\label{17a}\end{equation}
where ${\rm Det}  \Lambda _0/W_0 = t_b - t_a$.
Finally, setting to $g=1$ in (\ref{17a}) gives
 the required ratio of the functional determinants
\begin{equation}
 {\rm Det} K_0^{-1} \cdot K_1 =  \frac{{\rm Det}   \Lambda_g }{W_g}
  \bigg/ (t_b - t_a).
\label{18a}\end{equation}
In a time-sliced quantum mechanical path integral,
the determinant of $K_0$ is finite and has the value \cite{1}
\begin{equation}
  {\rm Det} K_0 = t_b - t_a,
\label{19a}\end{equation}
 so that we obtain
\begin{equation}
 {\rm Det} K_1 = \frac{[\eta_1(t_a )\xi_1(t_b) - \eta_1(t_b) \xi_1(t_a)]}{W_1},
\label{20a}\end{equation}
 which coincides
with  Gel'fand-Yaglom's formula (see Section 2.7 in \cite{1}).

For
 a  harmonic oscillator with a
time-dependent frequency $ \Omega (t)$ it is convenient
 to relate the set of two independent solutions $\eta_g (t)$ and $\xi_g (t)$
of Eq.~(\ref{9a}) at $g=1$, for which we omit the subscripts $g$,
 to the classical path $x_{\rm cl}(t) = x_a \xi (t) +
x_b \eta (t)$ satisfying the endpoint conditions $x_{\rm cl}(t_a) = x_a$
  and $x_{\rm cl} (t_b) = x_b$.
Since  this construction satisfies
$\eta_a = \xi_b = 0$,
  $\eta_b = \xi_a = 1$
 and $W= \dot  \xi_b = - \dot \eta_a$,  the explicit
 solution being
\begin{eqnarray}
  \xi (t) & = & \frac{\partial x_{\rm cl} (t) }{\partial x_a}  =
		\frac{p(t) p_b \sin  \omega _0 (q_b - q)}{p_a p_b \sin
                    \omega _0 (q_b  - q_a)}, \nonumber \\
  \eta (t) & = & \frac{\partial x_{\rm cl} (t)}{ \partial x_b}
		= \frac{p (t) p_a \sin  \omega _0 (q-q_a)}
                 {p_a p_b \sin  \omega _0 (q_b -  q_a) },
\label{14}\end{eqnarray}
They are
parametrized by two functions $q(t)$ and $p(t)$
satisfying
 the constraint
\begin{equation}
   \omega _0 \dot q p^2 = 1,
\label{15}\end{equation}
 where $ \omega _0$ is an arbitrary constant
frequency. The function $p(t)$ satisfies
the Ermakov-Pinney equation \cite{9}
\begin{equation}
  \ddot p +  \Omega ^2 (t) p - p^{-3} = 0.
\label{16}\end{equation}
 Inserting (\ref{14}) into  (\ref{18a}),
 we obtain for the
  harmonic oscillator with a
time-dependent frequency $ \Omega (t)$ the ratio
of functional determinants
\begin{equation}
  {\rm Det} K_0^{-1} K_1 = \frac{p_a p_b \sin  \omega _0 (q_b - q_a)}
         {(t_b - t_a)}.
\label{17}\end{equation}
where subscripts $a$ and $b$ indicate avaluation at $t=t_a$ and $t=t_b$,
respectively.
We check
 this representation
by  expressing the right-hand side
 in terms of the classical action $S_{\rm cl} (x)$.
With the same normalization
as in (\ref{19a}),
this yields
  the well-known one-dimensional Van-Vleck formula
\begin{equation}
  {\rm Det} K_1 = - M [\partial^2 S_{\rm cl} (x_a , x_b) / \partial x_a
      \partial x_b]^{-1}
\label{18}\end{equation}

To end this section we  note that the ratio (\ref{18a})
 can easily be extended to the stochastic case where the final position
of the trajectory $x(t)$ remains unspecified. To this end we
consider Eqs.~(\ref{14a}) and (\ref{10}) with a variable upper
time $t' \geq t \geq t_a$.  Then the eigenvalues
of the operator $K_0^{-1} K_1$ become  functions of $t'$
with a phase factor produced by each passage through a focal
point.
\\~\\
4.~Consider now  periodic (antiperiodic) boundary conditions
 $y (t_b) =
\pm y (t_a)$, $\dot y (t_b) = \pm \dot y (t_a)$
 for the eigenfunctions $y(t)$ of the operator $K_1
$ and
the for Green function $G^{p \atop a}(t,t')$:
\begin{eqnarray}
     G^{{\scriptscriptstyle p}\atop\raisebox{1ex}{$\scriptscriptstyle a$}}
(t_b,t') & = & \pm G^{{\scriptscriptstyle
p}\atop\raisebox{1ex}{$\scriptscriptstyle a$}} (t_a, t'),\nonumber \\
    \dot G^{{\scriptscriptstyle p}\atop\raisebox{1ex}{$\scriptscriptstyle a$}}
(t_b, t') & = & \pm \dot G ^{{\scriptscriptstyle
p}\atop\raisebox{1ex}{$\scriptscriptstyle a$}}
	 (t_a, t'),
\label{19}\end{eqnarray}
 where
$T = t_b - t_a$ is the period.
In both cases, the frequency $ \Omega  (t)$ and Dirac's $ \delta$-function
in Eq.~(\ref{4}) are also assumed to be periodic
(antiperiodic) with the same period.
 The general solution of Eq.~(\ref{4}) satisfying the boundary
conditions (\ref{19}) is constructed by adding to (\ref{7a})
an expression of the same  type
as before,
using the same
 homogeneous solutions
$\eta _g(t)$ and $\xi_g (t)$.
The result has the form
\begin{eqnarray}
 &&\!\!\!\!\!\!\!\!\!\!G_g^{{\scriptscriptstyle
p}\atop\raisebox{1ex}{$\scriptscriptstyle a$}} (t,t') = G_g (t,t')\nonumber \\
&& \mp
   \frac{\left[ f_g (t,t_a) \pm f_g (t_b,t) \right]
   \left[ f_g (t', t_a) \pm f_g (t_b, t')\right] }
   { \Delta ^{{\scriptscriptstyle p}\atop\raisebox{1ex}{$\scriptscriptstyle
a$}} \cdot f_g (t_a,t_b)}
\label{27a}\end{eqnarray}
with the condition
\begin{equation}
   \Delta
 ^{{\scriptscriptstyle p}\atop\raisebox{1ex}{$\scriptscriptstyle a$}}
 = \frac{{\rm Det}  \bar\Lambda
 ^{{\scriptscriptstyle p}\atop\raisebox{1ex}{$\scriptscriptstyle a$}}
 {}_{\!\!\!\!g}}{W_g}
    \neq 0,
\label{28a}\end{equation}
 where $\bar \Lambda ^{{\scriptscriptstyle
p}\atop\raisebox{1ex}{$\scriptscriptstyle a$}}$ are
now the  $(2 \times 2)$-constant matrices
\begin{equation}
   \bar\Lambda ^{{\scriptscriptstyle p}\atop\raisebox{1ex}{$\scriptscriptstyle
a$}} = \left(
\begin{array}{ll}
 (\eta_b \mp \eta _a) & ( \xi_b \mp \xi_a)\\
 (\dot \eta_b \mp \dot \eta_a) & ( \dot \xi_b \mp \dot \xi_a),
\end{array}\right)
\label{29a}\end{equation}
evaluated at $g \neq 1$.
 In analogy to Eq.~(\ref{16a}) we now find the
 formula
\begin{equation}
{\rm Tr}  \,\Omega ^2 (t) G
 ^{{\scriptscriptstyle p}\atop\raisebox{1ex}{$\scriptscriptstyle a$}}
 {}_{\!\!\!\!g}(t,t') =
 - \partial_g \log \left(\frac{{\rm Det}
\bar \Lambda
 ^{{\scriptscriptstyle p}\atop\raisebox{1ex}{$\scriptscriptstyle a$}}
 {}_{\!\!\!\!g}}{W_g}\right).
\label{30a}\end{equation}
 Substituting  this into (\ref{5})
  and setting $g = 1$, we obtain  the ratio
 of the functional determinants
 for periodic boundary conditions
%
 %%%%%%%%%%%%
%
\begin{equation}
{\rm Det} \tilde K^{-1} \cdot K_1 =
    \frac{{\rm Det}   \Lambda_1 ^p}{W_1}\bigg/
   4 \sin^2 \frac{ \omega _0 (t_b - t_a)}{2}
\label{31a}\end{equation}
Here Det $\tilde K={\rm Det} (-\partial _t^2- \omega_0^2)$
is the fluctuation determinant of the harmonic oscillator,
which in the same normalization as in (\ref{19a})
is equal to
\begin{equation}
 \mbox{Det}~ \tilde K = 4 \sin^2  \frac{\omega _0 (t_b - t_a)}{2},
\label{@}\end{equation}
 and thus the  formula
\begin{equation}
  {\rm Det} K_1 = \frac{(\eta_b- \eta_a) (\dot\xi_b -
   \dot\xi_a) - (\xi_b - \xi_a ) (\dot \eta_b - \dot \eta_a)}{W},
\label{33a}\end{equation}
the right-hand side being evaluated at $g=1$.
 For antiperiodic
boundary conditions, the analogous
 expressions are
\begin{equation}
 {\rm Det} \tilde K^{-1}_1 =
 \frac{{\rm Det}   \Lambda_1 ^a}{W_1}
\bigg/
   4 \cos^2 \frac{ \omega _0 (t_b -t_a)}{2}  ,
\label{34a}\end{equation}
\begin{equation}
{\rm Det} K_1 =  \frac{(\eta_b + \eta_a) (\dot\xi_b + \dot\xi_a)
       - (\xi_b + \xi_a)(\dot \eta_b + \dot\eta_a)}{W} .
\label{35a}\end{equation}
For a
  harmonic oscillator with a
time-dependent frequency $ \Omega (t)$,
 we  use again the representation
(\ref{14}) for $\xi(t)$ and $\eta(t)$
in terms of the functions $p(t)$ and $q(t)$, which in
addition to
 (\ref{15})
 and (\ref{16})  have the following properties: the
function $p(t)$ is periodic and even
\begin{equation}
 p (t + T) = p (t), ~~p(-t)= p(t)
\label{28}\end{equation}
 so that $p_b = p_a$, whereas the function $q(t)$ satisfies
\begin{equation}
  q(t+ T) = q(t) + q_b, ~~q_a = 0,
\label{29}\end{equation}
 where $T \equiv (t_b - t_a)$. Inserting now the solutions
 (\ref{14})  into  (\ref{31a})
and (\ref{34a}), we find the ratio of functional determinants for
a  harmonic oscillator with a
time-dependent frequency $ \Omega (t)$
with periodic boundary
conditions
\begin{equation}
  {\rm Det} \tilde K^{-1} K_1 = 4 \sin^2 \frac{ \omega _0 q_b}{2}
    \big/  4 \sin ^2 \frac{ \omega _0 t}{2},
\label{30}\end{equation}
 and with antiperiodic boundary conditions
\begin{equation}
  {\rm Det} \tilde K^{-1} K_1 = 4 \cos^2 \frac{ \omega _0q_b}{2}\big/
 4 \cos^2 \frac{ \omega _0 t}{2}.
\label{31}\end{equation}
Note that only
formula
 (\ref{17}) for the Dirichlet boundary condition
has been known in the literature (see \cite{cheng}--\cite{8}).
The periodic and antiperiodic
  formulas (\ref{30}) and (\ref{31})
are new, although they have had predecssors
on the lattice \cite{pre}.
Moreover, our new derivation
has the advantage of employing only
 Wronski's simple
construction  method for Green functions.
The general expressions for the functional
determinants  (\ref{20a}), (\ref{33a}) and (\ref{35a}) are
form-invariant under an arbitrary changes
$(\eta, \xi) \rightarrow (\tilde \eta, \tilde \xi)$
of the basic set $\eta(t)$ and $\xi (t)$ of two independent
solutions of the homogeneous equation (\ref{9a}).
\\~\\
5.~
%Consider now the situation where the operator $K_1$
%has a zero mode.
Contrary to the case of a
  harmonic oscillator with a
time-dependent frequency $ \Omega (t)$,
consider now the situation where the operator $K_1$ has a zero mode.
In this case we may assume  the frequency $ \Omega (t)$
in Eq.~(\ref{1}) the special form $ \Omega ^2(t) =
V''(x_{\rm cl}(t))/M$ with a potential $V(x)$,
 allowing
 reflecting the translation invariance of the theory with Lagrangian (\ref{1})
along the time axis
\cite{pre1}.
Let $\xi (t)$
be the corresponding eigenfunction
 satisfy the condition
 $\xi_a = 0$ as well as $\xi_b = 0$.
 %As  mentioned
%above, the  Wronskian $W$ of two independent
% solutions $\xi(t)$ and $\eta(t)$ satisfying  $\eta_n = \xi_b = 0$
%is now equal to zero identically
% $ W = \dot \xi_a \eta_a - \dot \eta_a \xi_a \equiv 0$,
% making Eq.~(\ref{7})
%undefined an
%   the ratio of determinants
%(\ref{13}) vanish.
%%%%%%%%%%%%%5
As mentioned above, the condition (\ref{12a}) is now violated,
making Eq.~(\ref{11a}) undefined, and it
 is impossible
to construct two independent solution $\xi(t)$ and $\eta (t)$
since their Wronski determinant
 would be equal to zero indentically $W = \eta_a \dot\xi_a - \dot\eta_a \xi_a
\equiv 0$ due to the boundary conditions (\ref{6}).
Since the Wronski construction
is not  applicable we replace
\begin{equation}
  \xi_b = 0,~~~\xi_a = 0
\label{32}\end{equation}
 by the regularized conditions
\begin{equation}
 \xi_b^\varepsilon = 0, ~~~\xi_a^\varepsilon
      = \varepsilon.
\label{33}\end{equation}
These
 do not require a
 new calculation of the determinant
(\ref{20a}),
and
  we find immediately
\begin{equation}
{\rm Det} K_1^\varepsilon = - \frac{\varepsilon }{\dot \xi_b}
    \rightarrow 0,
\label{34}\end{equation}
 in the limit $\varepsilon \rightarrow 0$.
We therefore remove the zero mode from
the determinant using
the standard method \cite{pre2}.
The regularized determinant  is defined by
\begin{equation}
  {\rm Det} K_1  ^R = \lim_{\varepsilon \rightarrow 0}
   \frac{{\rm Det} K_1^\varepsilon}{ \lambda ^\varepsilon},
\label{35}\end{equation}
 where $ \lambda ^\varepsilon$ is the eigenvalue associated with the
eigenfunction $\xi^\varepsilon (t)$.
\begin{equation}
  K_1 \xi^\varepsilon =  \lambda ^ \varepsilon \xi^\varepsilon,
\label{36}\end{equation}
with the limits
  $\xi^\varepsilon \rightarrow \xi, ~ \lambda ^\varepsilon \rightarrow
 0$ for $\varepsilon \rightarrow 0$. To  first order in $\varepsilon $
 it follows from (\ref{36}) that
\begin{equation}
  \int^{t_b}_{t_a} dt \xi K_1 \xi^\varepsilon \approx
    \lambda ^\varepsilon \int^{t_b}_{t_a} dt \xi^2
 (t) \equiv  \lambda ^\varepsilon \langle \xi | \xi\rangle.
\label{37}\end{equation}
 Integrating  the  left-hand side
 by parts and taking into account the conditions
(\ref{32}) and (\ref{33}) gives
\begin{equation}
   \lambda ^\varepsilon = - \varepsilon
\frac{\dot \xi_a}{\langle \xi|\xi\rangle}.
\label{38}\end{equation}
Finally, substituting (\ref{38}) and (\ref{34}) into (\ref{35})
we obtain the functional determinant without zero mode
\begin{equation}
 {\rm Det} K_{1}^R  =
  \frac{<\xi|\xi>}{\dot \xi_a\dot \xi_b}.
\label{39}\end{equation}
For periodic (antiperiodic) boundary conditions $y (t_a) = \pm y
  (t_b), \dot y (t_a) = \pm \dot y (t_b)$,
the analogous formula is
\begin{equation}
 {\rm Det} K_{1}^R  =
  \frac{(\xi_b\mp\xi_a)<\xi|\xi>}{ \eta_a( \eta_a\dot \xi_a-\dot \eta_a
\xi_b)}.
\label{39b}\end{equation}
In the periodic case,
 formula  (\ref{39b}) is useful
 for  semiclassical
calculations of path integrals
processing nontrivial classical solutions
such as solitons or instantons \cite{1}.
  \\~\\
 {\bf Acknowledgement:}\\
We thank A.\ Pelster for discussions.
The work was supported by Deutsche  Forschungsgemeinschaft
under grant Kl~256/28-1
and the Russian Foundation of Fundamental
Research under grant 97-01-00745.


\begin{thebibliography}{11}
%
\bibitem{1}
 H.~Kleinert, {\em Path Integrals in Quantum  Mechanics, Statistics,
    and Polymer Physics\/} (2nd edition), World Scientific, Singapore, 1995.
\bibitem{2}
  I.M.~Gel'fand and A.M.~Yaglom, J.~Math.\ Phys.~{\bf 1}, 48 (1960).
%\bibitem{1a}
% See Section 2.12 in \cite{1}.
%\bibitem{3}
%  R.~Forman, Invent.\ Math.\ {\bf 88}, 447 (1987).
%\bibitem{5'}
%See Sect.~A.~5 in Ref.~\cite{1}.
%\bibitem{4}
%A.J.~McKane and M.B.~Tarlie, J.~Phys.~{\bf A 28},
%    6931 (1995).
%%%%%%%%%%%%5
\bibitem{3}
 S.\ Levit and U.\ Smilansky, Proc.\ Am.\ Math.\
 Soc.\ {\bf 65}, 299 (1977).
\bibitem{5'}
 R.\ Forman, Inv.\ Math.\ {\bf 88}, 447 (1987);\\
  Commun.\ Math.\ Phys.\ {\bf 147}, 485 (1992).
\bibitem{4}
 D.\ Burghelea, L.\ Friedlander and T.\ Kappeler,
 Commun.\ Math.\ Phys.\ {\bf 138}, 1 (1991);\\
 Int.\ Eq.\ Op.\ Th.\ {\bf 16}, 496 (1993);\\
 Proc.\ Am.\ Math.\ Soc.\ {\bf 123}, 3027 (1995).
\bibitem{5}
 M.\ Lesch and J.\ Tolksdorf, dg-ga/9707022v1 (1997).
%
\bibitem{7}
 D.B.\ Ray, Adv.\ Math.\ {\bf 4}, 109 (1970);\\
 D.B.\ Ray and I.M.\ Singer, Adv.\ Math.\ {\bf 7},
 145 (1971);\\
 Ann.\ Math.\ {\bf 98}, 154  (1973).
%%%%%%%%%%
\bibitem{cheng}
     B.K.~Cheng, J.\ Math.\ Phys.\ {\bf 25}, 1804 (1984);\\
    J.\ Math.\ Phys.\ {\bf 27}, 217 (1986).
\bibitem{khanlaw}
   D.C.\ Khandekar and S.V.\ Lawande, J.\ Math.\ Phys.\ {\bf 20},
   1870 (1979).
%
\bibitem{rezende}
    J.\ Rezende, J.\ Math.\ Phys.\ {\bf 25}, 3264 (1984).
%
\bibitem{8}
   D.C.\ Khandekar and S.V.\ Lawande, Phys.\ Rep.\ {\bf 137},
     115 (1986).
\bibitem{12}
 M.\ Kontsevich and S.\ Vishik, hep-th/9405035 (1994).
%
%\bibitem{10}
%     D.B.\ Ray and I.M.\ Singer, Adv.\ Math.\ {\bf 7},
%    145 (1971).
%\bibitem{11}
%     R.\ Seeley, Am.\ J.\ Math.\ {\bf 91}, 963 (1966).
\bibitem{9}
   A.K.\ Common, E.\ Hesameddini and M.\ Musette,
   J.\ Phys.\ {\bf A 29}, 6343 (1996).
%\bibitem{10}
%\bibitem{12}
% M.\ Kontsevich and S.\ Vishik, hep-th/9405035 (1994).
\bibitem{14}
 A.J.\ McKane and M.B.\ Tarlie,
 J.\ Phys.\ {\bf A28}, 6931 (1995).
\bibitem{pre}
See Section 2.12 of Ref.~\cite{1}.
\bibitem{pre1}
See Section 17.3 of Ref.~\cite{1}.
\bibitem{pre2}
See Section A.5 in \cite{1} and \cite{14}.

\end{thebibliography}
\end{document}